\documentclass[mathleft
]{an}
\usepackage{graphicx}
\usepackage{times}
\usepackage{multirow}

\begin{document}

\Pagespan{789}{}
\Yearpublication{2012}%
\Yearsubmission{2012}%
\Month{}%
\Volume{}%
\Issue{}%

\title{Constraints on stellar parameters of the SPB star HD182255\\ from complex asteroseismology}

\author{Walczak, P.\thanks{Corresponding author: \email{walczak@astro.uni.wroc.pl}\newline}
  Szewczuk, W.
\and  Daszy\'nska-Daszkiewicz, J.
}
\titlerunning{Instructions for authors}
\authorrunning{T.H.E. Editor \& G.H. Ostwriter}
\institute{Instytut Astronomiczny Uniwersytet Wroc{\l}awski,\\
ul. Kopernika 11, 51-622 Wroc{\l}aw, Poland}

\received{2012}
\accepted{2012}
\publonline{later}


\abstract{Seismic modeling of the Slowly Pulsating B-type star HD182255 yields strong constraints
on the radial orders of the two dominant modes. For these frequencies, we derive also the empirical values of
the complex nonadiabatic parameter $f$. These two seismic tools, i.e., pulsational frequencies and the associated values of $f$,
allow to test available opacity data, chemical composition and overshooting efficiency.}

\maketitle

\section{Introduction}
Slowly Pulsating B-type stars (SPB) pulsate in high order g-modes, which penetrate the deep interior of the star.
These modes could be useful in particular to determine the overshooting efficiency from the convective core. Because of
a very dense oscillation spectrum of high order g-modes, seismic modeling the SPB stars is quite challenging
if no regular patterns are found.
However, it appeared that fitting at least two frequencies reduces greatly possible combinations
of the radial orders. Our analysis of the SPB star HD74560 (Walczak, Szewczuk \& Daszy\'nska-Daszkiewicz 2012), has shown that
only two combinations of the radial orders are possible.

In our seismic modelling we include also the empirical values of the nonadiabatic complex $f$-parameter, defined as the ratio of the radiative flux change to the radial displacement at the level of the photosphere (Daszy\'nska-Daszkiewicz, Dziembowski \& Pamyatnykh 2003, 2005). The values of $f$
depend on the properties of the superficial layers and, in case of B-type stars, are sensitive to stellar opacities and metallicity.
Fitting simultaneously frequencies and the $f$-parameters (so called complex asteroseismology) can provide better constraints on the parameters of the model and theory.

In the next section, we give basic information about the star HD182255. Results of the mode degree identification are presented in Section~\ref{sec:mi}. In Section~\ref{sec:ro}, we determine the radial orders of dominant modes and section~\ref{sec:sm} contains description of complex seismic models of the star. Conclusions end the paper.

\section{HD182255}
HD182255 is a bright star in the constellation of Vulpecula with about 5 mag in the Johnson V filter. The spectral type of the star is B6 and the Hipparcos (ESA 1997) parallax amounts to about 8.30 mas. According to G{\l}ebocki \& Gnaci\'nski (2005) the projection of its rotational velocity is equal to $V\sin i=$30 km/s. Hube \& Aikman (1991) discovered that the star is in a binary system with the orbital period of about 367.26 days. A slightly larger period (367.76 days) was found by Dukes et al.~(2003). The later authors discovered also three other repeatable variations with periods of $\sim1$ day. Similar values have been derived by Mathias et al.~(2001). These variations were interpreted as pulsations in high order g-modes. Five pulsational frequencies were found by De Cat et al.~(2007) in data gained during photometric and spectroscopic campaign devoted to SPB stars (Aerts et al.~1999). All frequencies were present in the Geneva photometry while the two dominant ones were detected also in spectroscopy (De Cat et al.~2007).

In order to estimate the basic parameters of the star, in Fig.~\ref{fig:HRdn} we show its position in the HR diagram. The effective temperature, $\log{T_{\rm{eff}}}=4.149\pm0.012$, was taken from Niemczura (2003) who determined this value from the IUE (International Ultraviolet Explorer, Bogges et al.~1978) spectra. The luminosity, $\log{L/L_{\odot}}=2.457\pm0.087$, was derived from the Hipparcos parallax and bolometric correction from Code at al.~(1976). We took into account also the interstellar extinction in the form of $A_V=RE(B-V)$, where we adopted the standard value of $R=3.1$ and the reddening of $E(B-V)=0.002\pm0.011$ (Niemczura 2003).
The evolutionary tracks from the Zero Age Main Sequence (ZAMS) to the Terminal Age Main Sequence (TAMS) were calculated by using the Warsaw-New Jersey evolutionary code. We adopted the AGSS09 chemical mixture (Asplund et al.~2009) and the OPAL opacity tables (Iglesias \& Rogers 1996) supplemented by the Ferguson, Alexander \& Allard~(2005) opacities. Furthermore, we assumed the initial rotational velocity of $V_{\rm{rot}}=30$ km/s, initial hydrogen abundance $X=0.7$, metallicity parameter $Z=0.02$ and no overshooting from the convective core. The lines marked as $g_i$ will be described
in the next Section. The mass of HD182255 estimated from evolutionary tracks is about $4M_{\odot}$ and most probably the star is in the main sequence phase.

\section{Mode identification}
\subsection{The mode degree, $\ell$}
\label{sec:mi}

To determine the mode degree of the five detected frequencies of HD182255 (De Cat et al.~2007), we compared firstly the theoretical values of the photometric amplitude ratios and phase differences with the observational counterparts using the theoretical values of the nonadiabatic $f$-parameter. The theoretical photometric observables were computed for models located in the center and on the edges of the observational error box. In the next step, the mode degrees were derived together with the empirical values of the $f$-parameter. This method was applied only to the two dominant frequencies because in case of the B-type stars the empirical value of the $f$-parameter can only be derived for frequencies which are visible in both photometry and spectroscopy (Daszy\'nska-Daszkiewicz, Dziembowski \& Pamyatnykh 2005). The Geneva photometric system consists of seven filters and the response functions of the adjacent ones overlap. Therefore, we used only $UBV$ which can be treated as independent. The above two approaches need input from models of stellar atmospheres. We used the LTE models by Kurucz (2004) and the non-linear limb darkening law by Claret (2003).

In Table~\ref{tab:mi} we present a summary of our identification of $\ell$. In the first two columns we give frequencies and the amplitudes of the light variation in the Geneva $V$ filter, respectively. The most probable mode degrees derived with the first and second methods are given in the third and fourth columns, respectively. The first method relied on photometry while the second - on photometry and spectroscopy.

\begin{table}[h]
\begin{center}
 \caption{Pulsational frequencies of HD182255, the Geneva $V$ amplitudes
  and the most probable identification of $\ell$ derived with the theoretical and empirical values of the $f$-parameter.}
 \label{tab:mi}
 \begin{tabular}{cccc}
\hline\noalign{\smallskip}
Frequency & $A_V$ & \multicolumn{2}{c}{Mode degree, $\ell$}\\
          $[$c/d$]$              &  [mmag]      & theoretical $f$ &empirical $f$\\\hline
$\nu_1$=0.97185(3) & 12.4(1.1)&$\ell=1$   &$\ell=1,2$\\
    $\nu_2$=0.79225(4) & 13.4(1.0)&$\ell=1$   &$\ell=1,2$\\
    $\nu_3$=0.62526(7) &  8.3(1.0) &$\ell=1,2$ &-  \\
    $\nu_4$=1.12780(10) &  3.5(1.0) &$\ell=1$   &-  \\
    $\nu_5$=1.02884(11)&  2.4(1.1) &$\ell=1$   &-  \\\hline
 \end{tabular}
 \end{center}
\end{table}

As we can see, from the first and second method we obtained that the dominant frequencies, $\nu_1$ and $\nu_2$, are the dipole modes. The second method indicates also the quadrupole modes. The dipole modes are also the $\nu_4$ and $\nu_5$ frequencies. The only ambiguous result is for the $\nu_3$ frequency, which can be $\ell=1$ or 2.

Unfortunately, we do not have information about azimuthal orders, $m$. Identification of $m$ requires the analysis of line profile variations or taking into account rotational effects which are beyond the scope of this paper. In the further analysis, we assume that pulsational modes of HD182255 are axisymmetric, i.e. $m=0$.

\subsection{The radial order, $n$}
\label{sec:ro}

\begin{figure}[h]
             \begin{center}
             \includegraphics[width=8.3cm,height=7.cm,clip]{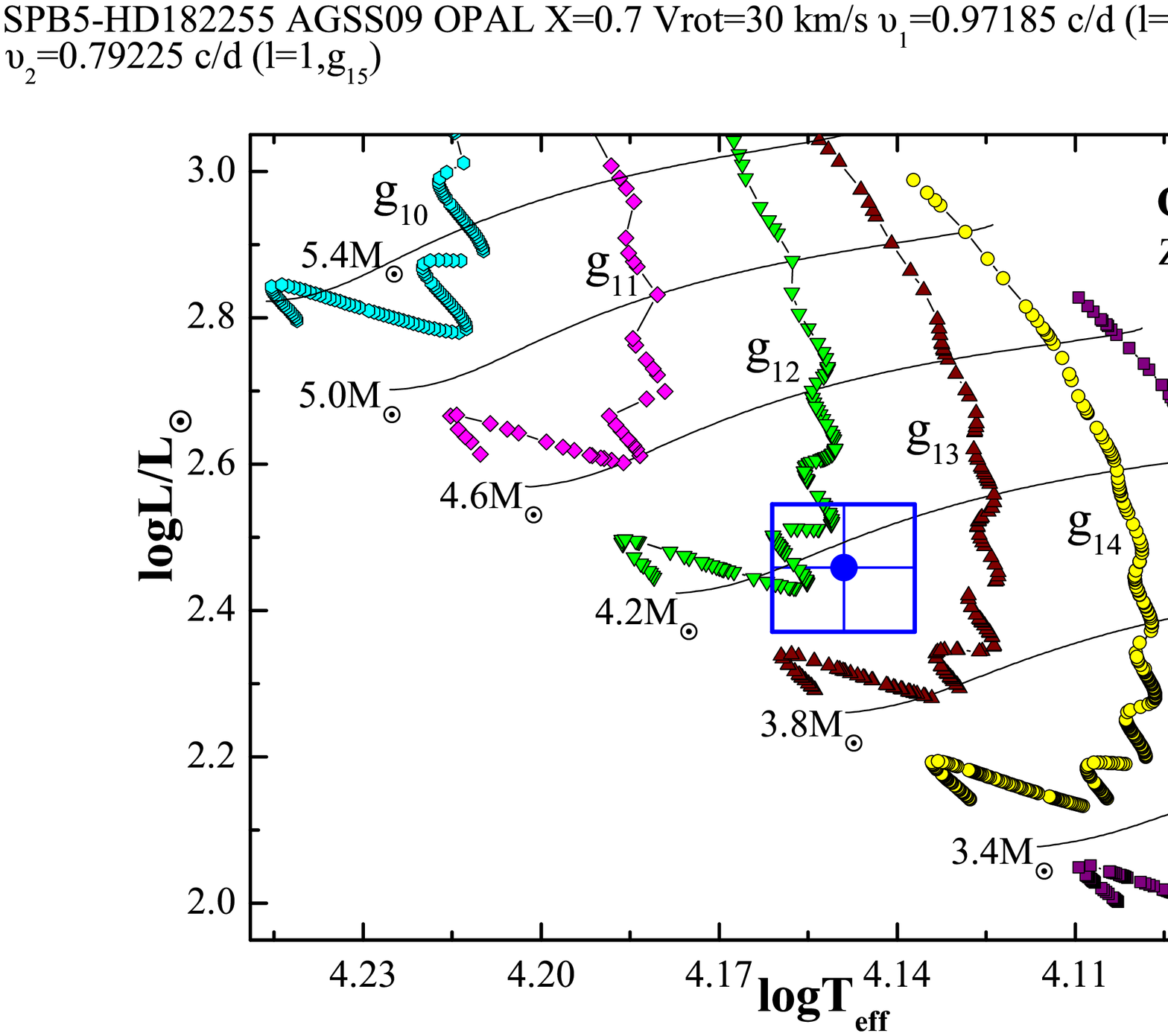}
             \caption{\small{The position of HD182255 in the HR diagram. Different symbols indicate models fitting the $\nu_1=0.97185$ c/d frequency assuming radial orders from $n=10$ to $n=15$. All models and evolutionary tracks were calculated with the OPAL opacities and metallicity $Z=0.02$. We did not include the overshooting effect.}}
             \label{fig:HRdn}
             \end{center}
\end{figure}

We started our modeling by fitting the dominant frequency $\nu_1=0.97185$ c/d assuming different radial orders, $n$.
Fig.~\ref{fig:HRdn} shows the position of models fitting $\nu_1$ in the HR diagram in cases when $n$ has a value from $10$ to $15$.
Each line is marked with the corresponding value of $n$.

As we can see there are a lot of models at different evolutionary stages as well as various radial orders allowed. In general, for a given radial order we have a more evolved star when the stellar mass increases. And we get more massive stars with decreasing radial order of the $\nu_1$ mode. These effects are easily explained by the facts that for high-order g-modes
the following rules apply:
\begin{itemize}
\item for a given mass the frequency decreases with increasing radial order,
\item for a given radial order, the frequency increases with decreasing stellar mass,
\item for a given radial order and mass, the frequency generally increases with time.
\end{itemize}
When we try to fit a constant value of a frequency assuming increasing radial order we have to reduce the mass. With smaller masses we get lower effective temperatures and luminosities And for a given radial order, the bigger the mass we choose, the more evolved a star becomes. However, we have to bear in mind, that there are exceptions to these rules connected with the avoided crossing which appears for non-radial modes and modifies the frequency evolution.

\begin{figure}[h]
             \begin{center}
             \includegraphics[width=8.3cm,height=5.cm,clip]{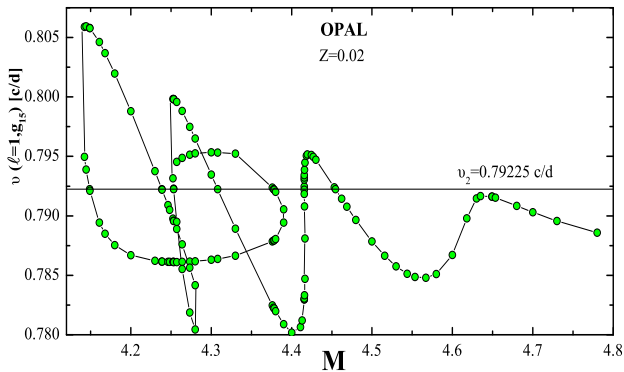}
             \caption{\small{Frequency of the $\ell=1$, g$_{15}$ mode as a function of the mass for models fitting $\nu_1$ (assuming that  $\nu_1$ is the $\ell=1$, g$_{12}$ mode). The results are for metallicity of $Z=0.02$. A horizontal line indicates the value of the $\nu_2$ frequency.}}
             \label{fig:M-ni2}
             \end{center}
\end{figure}

In the next step, we plotted these models in the frequency $vs.$ mass diagrams. We chose the modes with frequencies closest to the observational value of $\nu_2=0.79225$ c/d. An example of such diagram in shown in Fig.~\ref{fig:M-ni2}, where we plotted the frequency of the $\ell=1$, g$_{15}$ mode as a function of the mass. All these models fit the $\nu_1$ frequency under assumption that it is the $\ell=1$, g$_{12}$ mode. The horizontal line indicates a value of the $\nu_2$ frequency. As we can see, the frequency dependence of the mass is very complicated, but there are models fitting simultaneously the $\nu_1$ and $\nu_2$ frequencies. These models have the masses of about $4.15M_{\odot}$, $4.23M_{\odot}$, $4.27M_{\odot}$, $4.31M_{\odot}$, $4.375M_{\odot}$, $4.42M_{\odot}$ and $4.45M_{\odot}$.

In Fig.~\ref{fig:HRdn} we can easily see, that for $Z=0.02$ some models calculated with the radial order $n=12$ fit quite well the observational effective temperature and luminosity. We searched for models fitting two frequencies assuming different radial orders of the $\nu_1$ frequency.
As a result we found that models fitting $\nu_1$ and $\nu_2$ and laying close to the observational error box (which means closer than the $3\sigma$ errors) exists only if $\nu_1$ is the $\ell=1$, g$_{12}$ mode and $\nu_2$ is the $\ell=1$, g$_{15}$ mode.

\section{Complex asteroseismic analysis}
\label{sec:sm}

\begin{figure}[h]
             \begin{center}
             \includegraphics[width=8.3cm,height=7.cm,clip]{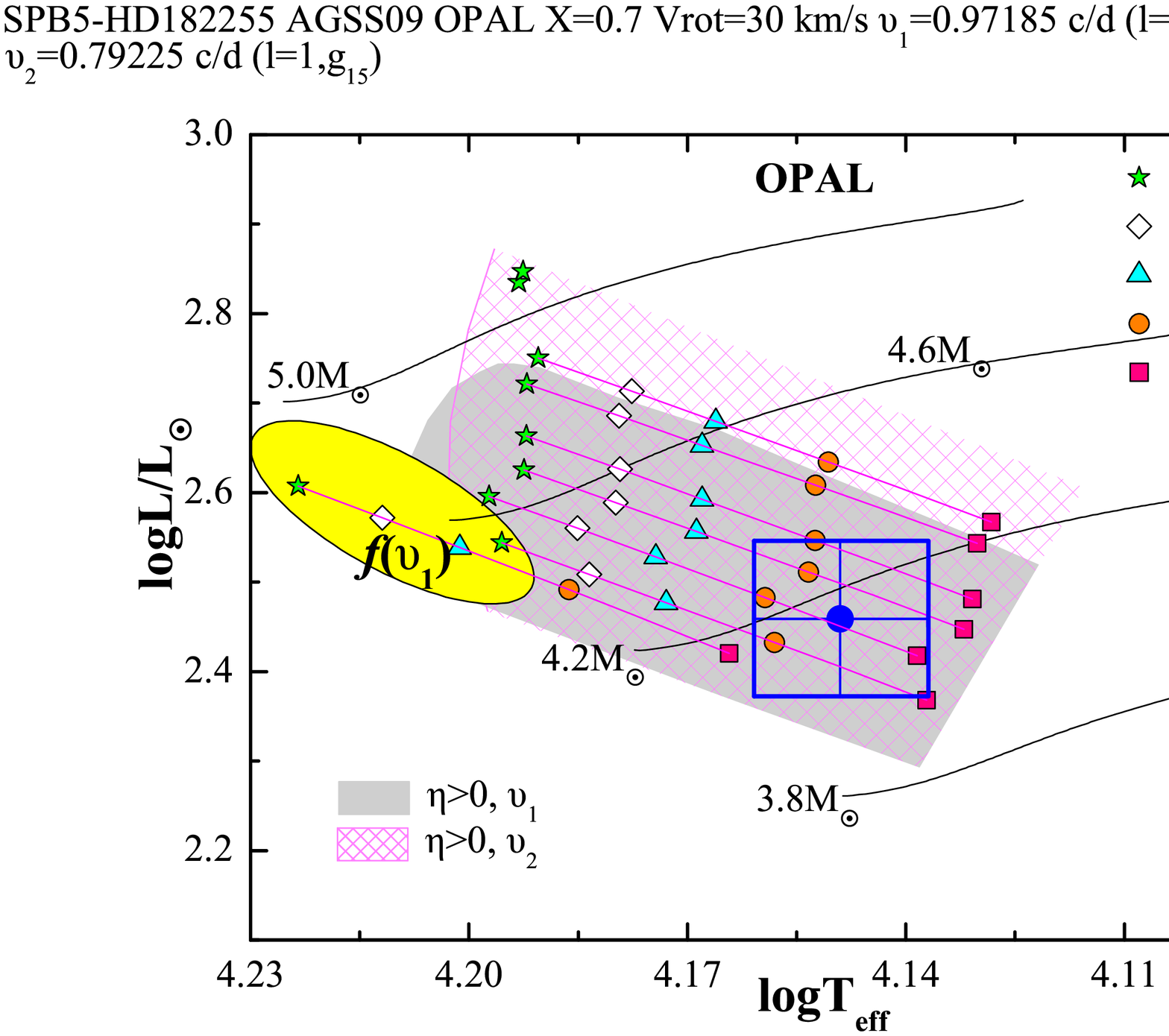}
             \caption{\small{The HR diagram with the location of the seismic models fitting the two dominant frequencies: $\nu_1$ ($\ell=1$, g$_{12}$) and $\nu_2$ ($\ell=1$, g$_{15}$). Symbols mark the values of metallicity, $Z$. Gray and hatched regions indicate domains where $\nu_1$ and $\nu_2$ are unstable, respectively. The area labeled as $f(\nu_1)$ is described in the text.}}
             \label{fig:HR-2nu}
             \end{center}
\end{figure}

In this section we consider only seismic models fitting simultaneously the two dominant frequencies: $\nu_1$ ($\ell=1$, g$_{12}$)
and $\nu_2$ ($\ell=1$, g$_{15}$). Such models are plotted in the HR diagram in Fig.~\ref{fig:HR-2nu}, where we show the metallicity effect. Along the lines connecting points there should also exist seismic models. However, between the lines there are no seismic models. The grey area indicates the models with the unstable $\ell=1$, g$_{12}$ mode while the hatched regions indicate models with the unstable $\ell=1$, g$_{15}$ mode. The area labeled as $f(\nu_1)$ will be discussed later.

The effective temperature of seismic models is highly sensitive to the metallicity parameter. In general, a smaller value of $Z$ gives higher effective temperature. We can also notice that for a given metallicity we have rather a large spread in luminosity while the effective temperature is only slightly variable. In order to catch models inside of the error box we need quite a high metallicity parameter; from about $Z=0.02$ up to $0.025$. This is rather inconsistent with the metallicity derived by Niemczura (2003): $[m/H]=-0.30\pm0.31$ which corresponds to $Z=0.0098_{-0.0050}^{+0.010}$. All models lying inside of the error box have unstable modes corresponding to $\nu_1$ and $\nu_2$.

Because the two dominant frequencies were detected in both photometry and spectroscopy we were able to derive the empirical values of the non-adiabatic $f$-parameter and compare them with theoretical counterparts. Unfortunately, we did not fully succeed. We managed to find models fitting the $f$-parameter only for the $\nu_1$ mode and within the $3\sigma$ errors. The models lie inside of the area labeled as $f(\nu_1)$ in the HR diagram (Fig.~\ref{fig:HR-2nu}). It can be  easily seen, that they are far from the error box; they have too high effective temperature. The $f$-parameter indicates very young models, close to ZAMS, and metallicity lower than 0.015.

One possible explanation of this disagreement between theoretical and empirical $f$-parameters is connected with the data. The problem is that, the spectroscopic observations were made about two years before photometry. The value of the empirical $f$-parameter is sensitive to both photometric and spectroscopic amplitudes and phases. If the amplitudes or phases changed significantly between observations, the obtained value of $f$ would be incorrect. This is justified suspicious since some seasonal variations in amplitudes and phases in Str\"omgren filters were found by Duces et al.~(2003).

\begin{figure}[h]
             \begin{center}
             \includegraphics[width=8.3cm,height=5.cm,clip]{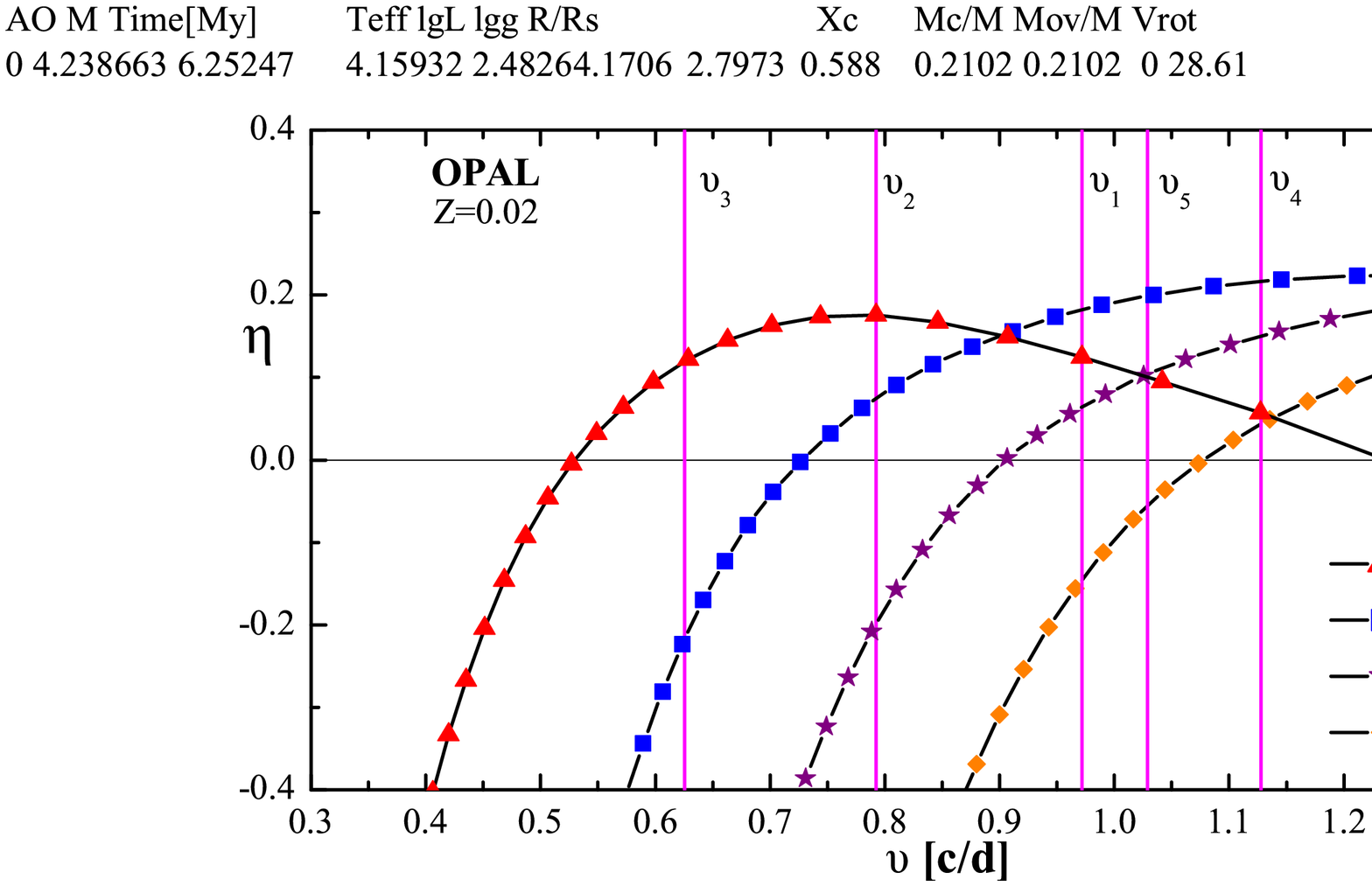}
             \caption{\small{Instability parameter as a function of frequency for modes with $\ell=1-4$. The chosen model is described in the text. The vertical lines indicate observational frequencies.}}
             \label{SPB5-HR-OP-OPAL-ni-eta}
             \end{center}
\end{figure}

We checked also whether our models fit other frequencies. It turned out that there is a model which reproduces quite accurately the $\nu_3$ and $\nu_4$ frequencies. Its parameters are: $M=4.239M_{\odot}$, $\log{T_{\rm{eff}}}=4.1593$, $\log{L/L_{\odot}}=2.4826$ and $Z=0.02$.
In Fig.~\ref{SPB5-HR-OP-OPAL-ni-eta} we plotted the instability parameter, $\eta$, of this model as a function of frequency for modes with $\ell=1-4$. Modes with $\eta>0$ are exited in the model. The vertical lines indicate observational frequencies of HD182255. The $\nu_3=0.62526$ c/d frequency can be both the dipole g$_{19}$ or quadrupole g$_{34}$ mode, but if it is the $\ell=2$ mode, it is stable ($\eta\sim-0.2$). The $\nu_4=1.12780$ c/d frequency is well reproduced by the dipole mode g$_{10}$ while the $\nu_5=1.02884$ c/d frequency does not have a theoretical counterpart. The rotational splitting of the $\ell=1$, g$_{15}$ mode is about 0.2 c/d (assuming rotational velocity of 30 km/s) so $\nu_5$ and $\nu_1$ could belong to the same triplet. It is interesting, that observational frequencies are inside of the instability domain of the dipole mode.

We took into account also overshooting from the convective core assuming $\alpha_{\rm{ov}}=0.1$, 0.2 and 0.4. Seismic models fitting the $\nu_1$ and $\nu_2$ frequencies changed slightly their position in the HR diagram. Moreover, we were not able to find models fitting other frequencies. Also the theoretical values of the non-adiabatic $f$-parameter did not fit better the empirical counterparts than in case of $\alpha_{\rm{ov}}=0$.

Quite similar models appeared with the OP opacity tables (Seaton 2005). The main difference was connected with the instability regions: with OP they were slightly larger. The quadrupole g$_{34}$ mode nearly fitting the $\nu_3$ frequency was almost unstable for $Z\ge0.02$. But still we could not find models fitting the empirical values of the $f$-parameter with effective temperatures and luminosities in agrement with the observational counterparts.

\section{Summary}

Fitting two observational frequencies allowed us to constrain radial orders of the pulsational modes. Seismic models close to the observational error box exist only for one combination of the radial orders: $\nu_1$ is the $\ell=1$, g$_{12}$ mode and $\nu_2$ is the $\ell=1$, g$_{15}$ mode. Including overshooting from the convective core did not improve our models. Moreover, with high values of $\alpha_{\rm{ov}}$ we could not find models fitting more than two frequencies. It seems, that effective overshooting for HD182255 is not necessary.

We did not succeed in fitting the empirical and theoretical values of the non-adiabatic $f$-parameter for both modes $\nu_1$ and $\nu_2$. The possible explanation is the inaccuracy in opacities on which the $f$-parameters strongly depend and/or lack of the simultaneous photometric and spectroscopic observations. Further observations are needed as well as identification of the azimuthal number $m$ for all pulsational modes.

\acknowledgements
The authors acknowledge partial financial support from the Polish
MNiSW grant No. N N203 379 636.


\end{document}